\documentclass[]{ceurart}

\usepackage{listings}
\usepackage[raggedrightboxes]{ragged2e}
\usepackage{amsmath}

\begin{document}
\copyrightyear{2023}
\copyrightclause{Copyright for this paper by its authors.
  Use permitted under Creative Commons License Attribution 4.0
  International (CC BY 4.0).}

\conference{CLEF 2023: Conference and Labs of the Evaluation Forum, September 18–21, 2023, Thessaloniki, Greece}
\title{Transfer Learning with Semi-Supervised Dataset Annotation for Birdcall Classification}
% \title[mode=sub]{Working Notes for Team DS@GT in the BirdCLEF 2023 Research Code Competition}

\author[1]{Anthony Miyaguchi}[
orcid=0000-0002-9165-8718,
email=acmiyaguchi@gatech.edu,
url=https://acmiyaguchi.me,
]
\cormark[1]

\author[1]{Nathan Zhong}[
email=nathanzhong@gatech.edu,
]

\author[1]{Murilo Gustineli}[
orcid=0009-0003-9818-496X,
email=murilogustineli@gatech.edu,
url=https://linkedin.com/in/murilo-gustineli,
]

\author[1]{Chris Hayduk}[
email=chayduk3@gatech.edu
]

\address[1]{Georgia Institute of Technology, North Ave NW, Atlanta, GA 30332}
\cortext[1]{Corresponding author.}

\begin{abstract}
We present working notes on transfer learning with semi-supervised dataset annotation for the BirdCLEF 2023 competition, focused on identifying African bird species in recorded soundscapes. 
Our approach utilizes existing off-the-shelf models, BirdNET and MixIT, to address representation and labeling challenges in the competition.
We explore the embedding space learned by BirdNET and propose a process to derive an annotated dataset for supervised learning. 
Our experiments involve various models and feature engineering approaches to maximize performance on the competition leaderboard. 
The results demonstrate the effectiveness of our approach in classifying bird species and highlight the potential of transfer learning and semi-supervised dataset annotation in similar tasks.
\end{abstract}

%% Keywords. Separate the keywords with commas.

\maketitle

\section{Introduction}

The BirdCLEF 2023 competition \cite{birdclef2023} focuses on classifying bird species in 10-minute-long soundscapes recorded in various parts of Africa as part of the LifeCLEF lab \cite{lifeclef2023}.
We need to label each 5-second interval in the test soundscape with the probability that each 264 target species is present.
The training dataset comprises 16,941 tracks spanning 192 hours of audio tagged by the label of the target species present.
While we have metadata about the species in the tracks, we do not have concrete labels on which audio sections contain calls.
The dataset also poses significant challenges given the species distribution, with many having only one or two examples.

We focus our efforts on utilizing existing off-the-shelf models as the basis for our classification models.
We hypothesize that we can reduce environmental noise and cross-talk using a sound-separation model and re-purpose knowledge in the embedding space learned by a domain-specific convolution neural network.
These models can address some of the most significant impediments in the competition, including the need for labeled data through semi-supervision.

\section{Embedding Space and Transfer Learning}

\begin{figure}[ht]
\centering
\includegraphics[width=\textwidth]{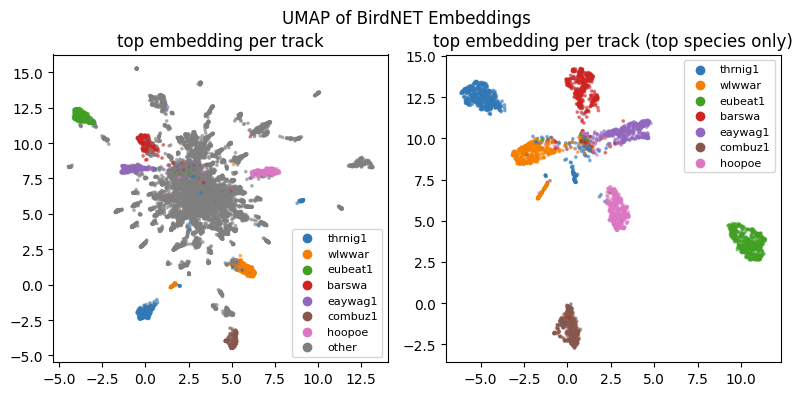}
\caption{
We demonstrate the clustering properties of the BirdNET embeddings by projecting them into $\mathcal{R}^{2}$ via UMAP \cite{mcinnes2020umap}
The projection preserves Euclidean distance in 2D space.
We take the embedding token across each track with the most significant probability across the BirdNET prediction vector and assign it a positive label.
The left plot shows clustering across the seven most common species in the training dataset.
The right plot demonstrates a clear separation between the seven most common species.
}
\label{fig:birdnet_embeddings}
\end{figure}

BirdNET \cite{kahl2021birdnet} classifies 48kHz 3-second audio clips into 3337 classes using a convolutional neural network trained on scaled spectrograms computed by the short-time Fourier transform.
The classes are primarily composed of bird species but include non-bird classes such as environmental noise or human voices.
We obtain embedding tokens by taking the values at the second-to-last layer of the model, preceding a fully connected logit layer.
The embedding maps the audio in the time domain into a vector space $\mathcal{R}^{320}$ that roughly preserves distances between points.
We use embedding tokens as features in a supervised machine-learning model to take advantage of the compact representation of the audio data.
We show a clear separation between the embedding tokens by running them through a dimensional reduction technique in figure \ref{fig:birdnet_embeddings}.
Clustering supports a hypothesis that we effectively utilize representation learned by BirdNET in new contexts.

\section{Semi-Supervised Dataset Annotation}

\begin{figure}[ht]
\centering
\includegraphics[width=\textwidth]{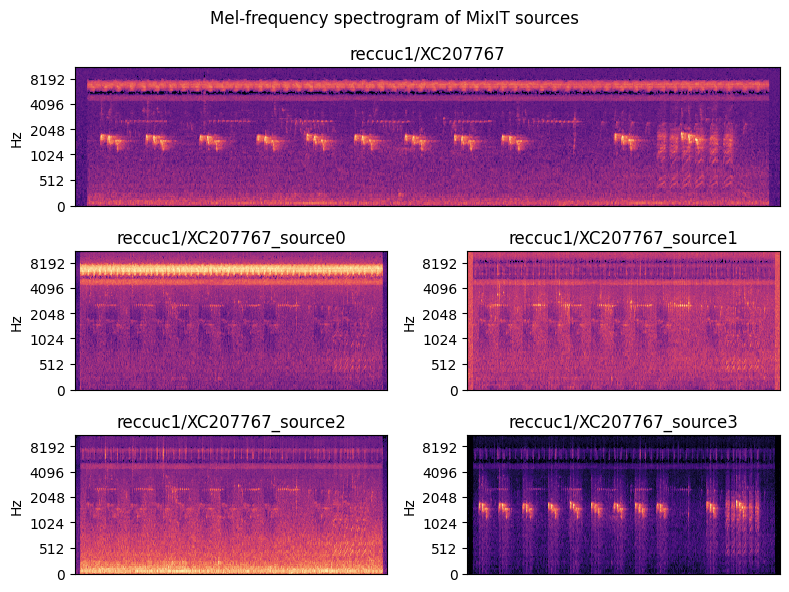}
\caption{
Bird-MixIT has improved the precision of downstream classifiers in experimental settings.
We demonstrate separation across entities in the mel-spectrogram of track \texttt{XC207767} containing a Red-chested Cuckoo.
We observe separating three sound signatures into sources 0, 1, and 3.
Source 2 is an amalgamation of the other sound signatures, an artifact of the model separating into four channels.
An automated process should choose source 3 containing the species of interest.
}
\label{fig:mixit}
\end{figure}

\begin{figure}[ht]
\centering
\includegraphics[width=\textwidth]{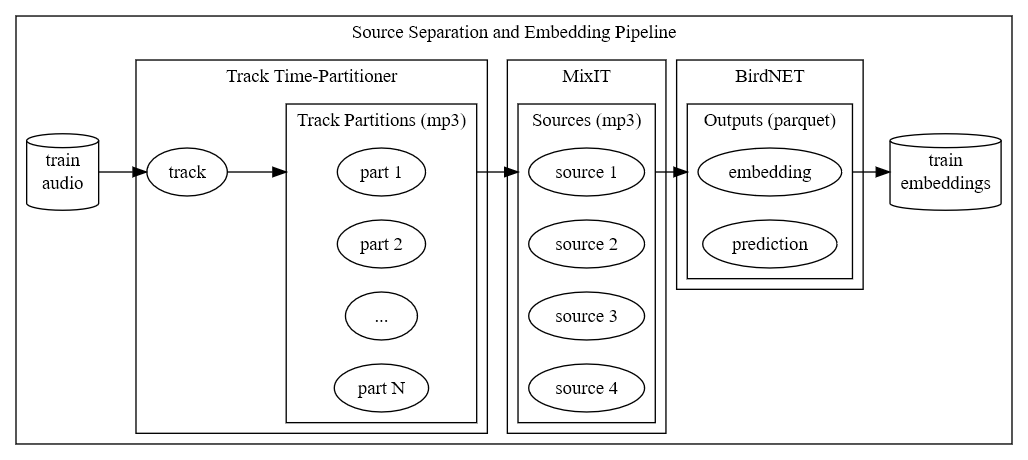}
\caption{
We use Luigi to coordinate a processing pipeline spanning days on an n2-standard-16 compute instance.
We prevent processing skew across workers by recursively training audio.
The audio is then source separated and embedded, resulting in a parquet file per audio chunk.
We consolidate the parquet files into the final dataset.
}
\label{fig:train-pipeline}
\end{figure}

We propose a process to derive an annotated dataset to fit data to traditional supervised learning algorithms.
First, we chunk audio within the training examples so that no track lasts 3 minutes by recursively splitting the tracks until they are smaller than our threshold, padded to the nearest third second with additive white noise.
Chunking the audio solves the problem of batch processing skew introduced by several examples that are longer than 30 minutes.
We assume the upper bound on the track length is sufficient to model temporal dependencies.
We use Bird-MixIT \cite{denton2022improving} to isolate environmental noise and bird vocalizations.
We then process each of the tracks using BirdNET to extract an embedding vector in $\mathbb{R}^{320}$ and a prediction logit vector in $\mathbb{R}^{3337}$ for every 3-second interval over a 1-second sliding window.
Each interval is labeled with top prediction labels and the energy of the original track.
We save the results from each track to disk and consolidate them into a parquet dataset.
See table \ref{tab:example-emb-v4} for an example row of this process.
We only have to pay for the expensive process of running TensorFlow models once by processing the audio training examples before fitting models.

We had four versions of the embedding dataset (\texttt{emb}).
In \texttt{emb\_v1}, we encountered missing entries and Docker permission problems. 
We fixed this in \texttt{emb\_v2} but incorrectly mapped input to the wrong model layer. 
Our first usable dataset was \texttt{emb\_v3}, which fixed previous issues and addressed the skew of long tracks by recursively chunking them.
\texttt{emb\_v4} includes all possible 3-second intervals at a 1-second resolution.
With the embedding and prediction vectors in a dataset, we apply a series of heuristics and feature engineering for our final models.

\begin{table}[ht]
\begin{tabular}{|l|l|}
\hline
\textbf{column name} & \textbf{value}                                                                         \\ \hline
species              & grecor                                                                                 \\ \hline
track\_stem          & XC629875\_part003                                                                      \\ \hline
track\_type          & source0                                                                                \\ \hline
track\_name          & grecor/XC629875\_part003\_source0.mp3                                                  \\ \hline
embedding            & {[}0.6731137633323669, 1.1389738321304321, 0.6284520626068115, ...     \\ \hline
prediction\_vec      & {[}-8.725186347961426, -7.3204827308654785, -9.82101821899414, ...     \\ \hline
predictions          & {[}\{0, 3026, Tetrastes bonasia\_Hazel Grouse, hazgro1, 0.02235649898648262\}, \{1,... \\ \hline
start\_time          & 75                                                                                     \\ \hline
energy               & 0.01598571054637432                                                                    \\ \hline
\end{tabular}
\caption{
Example output row from the embedding dataset (\texttt{emb\_v4}).
The \texttt{emb\_3} dataset is about 20 GB, and the \texttt{emb\_v4} dataset is about 60 GB.
}
\label{tab:example-emb-v4}
\end{table}

\section{Implementation and Workflow}

We split our workflow into training and inference.
Our training pipeline runs on the Google Cloud Platform (GCP), while inference runs in a Kaggle notebook optimized for offline usage.

We implement a shared Python package on GitHub. \footnote{Implementation at \
\href{https://github.com/dsgt-birdclef/birdclef-2023}{github.com/dsgt-birdclef/birdclef-2023}}
The package defines environment dependencies to run the training and inference workflows.
It contains helper PySpark code \cite{armbrust2015spark}, wrappers around BirdNET, and utilities for manipulating audio samples into matrices representing sliding windows.
We also define a workflow package containing Luigi \cite{Luigi2023} scripts that implement dataset processing and annotation.
We build Docker images for BirdNet and MixIT and integrate modified versions of the canonical inference script into our data pipeline as per figure \ref{fig:train-pipeline}.

The inference pipeline is composed of three notebooks.
The package sync notebook downloads the shared Python package with all dependencies into a local directory.
The model sync notebook similarly downloads serialized models and weights from object storage.
The final inference notebook runs offline after attaching the package and model sync notebooks as data sources.
We read the soundscapes, split them into chunks, obtained BirdNET embeddings, and computed predictions submitted to the competition.

\section{Experiments}

We run experiments on the embedding dataset to maximize our performance on the public leaderboard.
The feature engineering process of embedding tokens and prediction logits is part of the model-fitting process.
We first focused on a baseline with minimal modifications to the embedding dataset and then worked toward overcoming the idiosyncrasies of the training dataset by more complex feature engineering.
The input interval of BirdNET and the output interval of the competition do not match, so we had to consider this difference.
The former expects 3-second intervals, while the latter expects 5-second intervals.
Our two main approaches are to aggregate the output of models that represent 3-second intervals and to aggregate input to represent 5-second intervals.

While we use simple measures such as macro-precision and accuracy to assess models against the derived dataset in hyperparameter searches, we note that they did not reflect performance against the public leaderboard.
The results of our derived datasets and models are summarized in table \ref{tab:dataset-summary} and table \ref{tab:model-summary}, respectively.

\begin{table}[ht]
\begin{minipage}[b]{0.495\textwidth}
\centering
\footnotesize
\begin{tabular}{|l|l|}
\hline
\textbf{column name} & \textbf{value}      \\ \hline
track\_stem          & XC116777            \\ \hline
track\_type          & source1             \\ \hline
start\_time          & 0                   \\ \hline
primary\_label       & ratcis1             \\ \hline
metadata\_species    & {[}ratcis1{]}       \\ \hline
probability          & 0.348693 \\ \hline
embedding        & {[}0.64499, 0.45046, 0.36006, ... \\ \hline
next\_embedding  & {[}0.86384, 0.81126, 0.26452, ... \\ \hline
track\_embedding & {[}0.67923, 0.53318, 0.36459, ... \\ \hline
\end{tabular}
\end{minipage}
\hfill
\begin{minipage}[b]{0.495\textwidth}
\centering
\footnotesize
\begin{tabular}{|l|l|}
\hline
\textbf{column name} & \textbf{value} \\ \hline
track\_stem          & XC213642       \\ \hline
track\_type          & original       \\ \hline
start\_time          & 55             \\ \hline
species              & afmdov1        \\ \hline
embedding       & {[}1.76885, 0.83265, 0.49710, ... \\ \hline
prediction\_vec & {[}-14.44950, -12.09912, -15.69485, ... \\ \hline
\end{tabular}
\end{minipage}
\caption{
    Example output row from the \texttt{post} v6 dataset on the left and v7 on the right.
}
\end{table}

\begin{table}[h!]
\begin{tabular}{|l|p{0.5in}|p{4in}|}
\hline
\textbf{Dataset} & \textbf{Source} & \textbf{Description} \\ \hline
post v1 & emb v3 &
  Simplified dataset where embedding tokens come from the channel with the highest number of positive classifications against the baseline model. 
  Multi-label generated by averaging pairs and triplets of embedding tokens together. 
  \\ \hline
post v2 &
  emb v4 &
  Tokens are now the average of the first and third tokens of each 5-second interval. We generate current, next, and track embeddings using only source-separated tracks. Tokens are multi-labeled by the primary and secondary species associated with the track. \\ \hline
post v3 & post v2 &
  Augments above but the top-20 tokens in each species are averaged against random no-call tokens to simplify train-test splits. 
  \\ \hline
post v4 & emb v4 &
  Same methodology as post v2 and v3. 
  Multi-label is generated by confident baseline predictions filtered by plausible primary and secondary metadata labels. 
  \\ \hline
post v5 & emb v4 &
  Same as post v4, but it fixes a modulo bug in previous averaged-token datasets. 
  \\ \hline
post v6 & emb v4 &
  Drops notion of multi-label prediction. 
  It uses the original track and the best source-separated track to increase the number of training examples. 
  \\ \hline
post v7 & emb v4 &
  Adds logic to assign the primary label and no-call labels. 
  It uses concatenation instead of interpolation for a 5-second interval token and includes the prediction logits for the current interval. 
  \\ \hline
\end{tabular}
\caption{
An overview of changes in the post-processed dataset (\texttt{post}).
}
\label{tab:dataset-summary}

\begin{tabular}{|p{0.75in}|l|p{2.25in}|p{0.5in}|p{0.5in}|}
\hline
\textbf{Model}         & \textbf{Dataset}         & \textbf{Description}                                     & \textbf{Public Score} & \textbf{Private Score} \\ \hline
Logistic Regression & emb v3 & Baseline & 0.78541 & 0.68369 \\ \hline
MLP                    & emb v3         & Baseline       & 0.74014                     & 0.62283                      \\ \hline
XGBoost &
  post v1 &
  Multi-label one-vs-rest strategy. Interpolated token pairs and triplets. &
  0.79068 &
  0.68181 \\ \hline
XGBoost                & post v1 & Same as above, but weighted samples and native multi-label training. & 0.78829                     & 0.68053                      \\ \hline
XGBoost                & post v5 & Current interpolated-token.                       & 0.7692                      & 0.65937                      \\ \hline
XGBoost                & post v5 & Current and track interpolated-token.             & 0.7489                      & 0.63059                      \\ \hline
XGBoost                & post v3 & Current, next, and track interpolated-token.  & 0.75049                     & 0.63877                      \\ \hline
XGBoost                & post v3 & Current and next interpolated-token.              & 0.76484                     & 0.65346                      \\ \hline
XGBoost                & post v7 & Current concatenated token.               & 0.75997                     & 0.64414                      \\ \hline
XGBoost &
  post v4 &
  Ensemble of best logistic regression and boost model. &
  0.75091 &
  0.64242 \\ \hline
Complement Naive Bayes & post v7 & Current prediction logit softmax vector.        & 0.71093                     & 0.59652                      \\ \hline

\end{tabular}
\caption{A summary of experiments and their associated datasets.}
\label{tab:model-summary}
\end{table}

\subsection{Baseline Model}

The baseline model uses embedding tokens taken from the isolated track source with the highest energy, assuming that the loudest voice in the track is associated with the primary label.
We label the tokens according to the primary label if the max probability of the prediction vector exceeds a threshold, e.g., 0.5; otherwise, we label the token as "no-call".
We fit the data to logistic regression, multi-layer perceptron (MLP), support vector machine (SVM), and gradient-boosted decision tree (GBDT) classifiers.
We use scikit-learn \cite{scikit-learn} for the first three of these models and the scikit-learn compatible interface against XGBoost \cite{chen2016xgboost} for the latter.
When applicable, we perform a hyper-parameter search using sci-kit-optimize \cite{head2018skopt}, which performs sequential optimization using Bayesian methods.

We hypothesize that classes cluster together in the low-dimensional embedding learned by BirdNET and that linear models can effectively learn to discriminate between new classes.
Our baseline logistic regression classifier trained on the embedding tokens reaches a public/private score of 0.78/0.68, notably better than the starter Kaggle notebook using the Google Research Bird Vocalization Classifier with a score of 0.72/0.61. 
We find that SVM and GBDT are comparable to the logistic regression and that MLP models require significant tuning to reach good performance.

GBDT via XGBoost is our preferred model because it trains quickly with a GPU with relatively high predictive performance.
While logistic regression has fewer parameters and performs just as well, training can be slow as the number of examples increases.
In table \ref{tab:model-fit-time}, logistic regression takes 12x as long to train as XGBoost with GPU-based histogram binning.

\begin{table}[h]
\begin{tabular}{|l|l|l|l|}
\hline
Model                                                 & GPU & Fit time    & Predict time \\ \hline
Logistic Regression (Newton-Cholesky) & No  & 59 min 17 s & 1.5 s          \\ \hline
SVC                                                   & No  & 90 min +    & -              \\ \hline
MLP                                                   & No  & 4 min 14 s  & 3.5 s          \\ \hline
XGBoost (hist)                                        & No  & 48 min 20 s & 14.4 s         \\ \hline
XGBoost (gpu\_hist)                                   & Yes & 5 min       & 15.3 s         \\ \hline
ComplementNB                                          & No  & 4.2 s       & 1.5 s          \\ \hline
\end{tabular}
\caption{
A comparison between fit and predict the time for various models fit on a GCP n1-standard-4 compute instance with a Telsa T4 GPU. 
We fit the post-v7 dataset, which has 255,372 rows.
}
\label{tab:model-fit-time}
\end{table}

\subsubsection{Baseline Binary No-call Model}

We explore and analyze the performance of a binary classifier to further our understanding of the embedding space.
While we do not use this model directly in the competition, it helps quantify the quality of our automated labeling process.

We construct a binary dataset with positive and negative embedding samples.
Positive signifies the presence of a birdcall in the sample, while negative denotes its absence.
The distribution of positive and negative samples within the dataset is well-balanced, with 49.7\% being positive and 50.3\% being negative.
About half of the available training data is empty across original and sound-separated tracks.
A logistic regression classifier achieved an accuracy of 0.88 on this binary dataset.

\begin{figure}[h!]
\centering
\includegraphics[width=\textwidth]{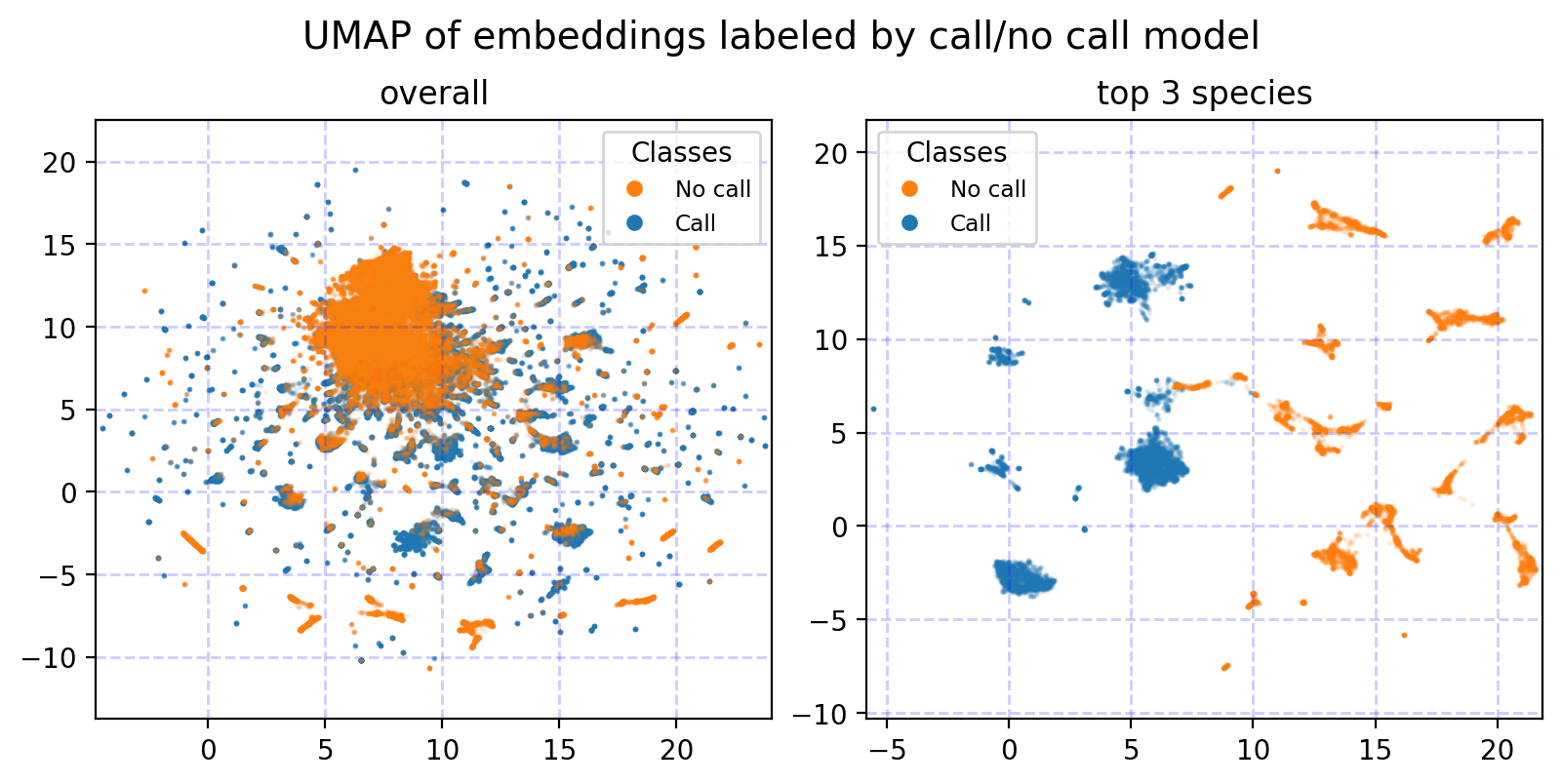}
\caption{
We demonstrate the clustering properties of the binary no-call embeddings by projecting them into $\mathcal{R}^{2}$ via UMAP. 
The clustering compares the binary dataset with the overall species versus the top-3 species.
}
\label{fig:birdnet_embeddings-umap-model}
\end{figure}

We create a second binary dataset from a subset of the first using the top three most common species, with 49.8\% of samples being positive and 50.2\% being negative.
A logistic regression classifier is trained and predicted 1.0 accuracy on the smaller dataset.
We hypothesize that this behavior is due to the large number of species in the dataset.
The distribution of some species is highly skewed, as shown in Figure \ref{fig:birdnet_embeddings-umap-model}.

Furthermore, we utilized the trained logistic regression classifier to make inferences a background audio embeddings based on the \texttt{freefield1010} dataset \cite{stowell2013open}.
The classifier predictions are probability scores indicating the presence or absence of birdcalls in each sample.
It predicted 77.3\% of the samples as having no birdcalls (no-call) and 22.7\% as having birdcalls (call). 
However, if the classifier's predictions were perfect, the no-call percentage should have been 1.0, as the entire dataset consisted only of background noise.

\subsection{Interpolated Embedding Models}
\label{sec:interpolated-emb model}

We build upon the baseline embedding model by interpolating embedding tokens to synthesize new examples and labels.
We assume that positive examples from each class tend to cluster together in the high-dimensional embedding space.
We also assume that embedding space takes on a Euclidean geometry or admits some approximation.
By interpolating examples, we hypothesize that the resulting coordinate lies between the clusters and, therefore, closer to the decision boundary of the sources.

We first use interpolation to address the alignment problem between the 3-second interval of BirdNET and the 5-second interval of the competition by generating a new feature.
In addition, we also construct a method that generates pairs and triplets of tokens sampled evenly across classes.
These interpolated tokens are assigned multiple labels used in a multi-label classifier.

We experiment with interpolation to add contextual information to each example.
For each token in a track from our dataset $v_{t}$, we generate the token that follows it directly in time $v_{t+1}$ and the token that represents the entire track $\sum_{i=0}^{n} v_i$.
We generate features by concatenating ($\oplus$) each of these tokens and evaluate performance relative to each other using the same set of labels.

\begin{align}
    \hat{y} &\sim M_1(v_t) \\
    \hat{y} &\sim M_2(v_t \oplus v_{t+1}) \\
    \hat{y} &\sim M_3(v_t \oplus \sum_{i=0}^{n} v_i) \\
    \hat{y} &\sim M_4(v_t \oplus v_{t+1} \oplus \sum_{i=0}^{n} v_i)
\end{align}

Interpolation may provide some value, in particular around multi-labeling.
We found that our model trained on a dataset does not perform worse than our baseline model while avoiding issues related to having a small number of training examples for the class.
Most models trained using a form of interpolated tokens resulted in lower leaderboard scores.
However, these models did not include interpolated pair and triplet tokens.

\subsection{Concatenated Embedding Model}

Another class of models that handles the time-interval discrepancy involves concatenating the embedding tokens.
We must train a classifier that accepts input in $\mathcal{R}^{2x320}$.
This model performs worse than the interpolated model.
The increased dimensional of the underlying feature also increases the model fit time, which leads us to skip augmented feature sets.

\subsection{Ensemble Embedding Model}

Our last embedding model uses the best models from our baseline and interpolated embedding experiments.
We train an XGBoost model trained on the outputs from the best classifiers.
This model is likely affected by the quality of the training dataset and the differences in embedding token semantics.

\subsection{Probability Logit Model}

Our final experiment uses the outputs of the final logit layer in the BirdNET model to determine a class' presence directly.
We generate a probability vector by taking a softmax of the logit layer.
We ran into out-of-memory issues on our GPUs fitting XGBoost models using the probability vector in $\mathcal{R}^{3337}$ and found the scikit-learn logistic regression implementation needed to be faster.
We fit the data using a Complement Naive Bayes classifier instead of a GBDT or logistic regression classifier.
Naive Bayes assumption works well in this problem, where each feature is treated independently toward the classification goal.
It is also swift because it simply computes counts over features.
We use the Complement Naive Bayes model to address the heavy skew in the class distribution but also find comparable performance across this family of classifiers.
The predictive performance is far worse than the baseline, with a public/private score of 0.71/0.59.

\section{Discussion}

\subsection{Semi-Supervised Annotation Quality}

The data quality is an aspect of our training dataset that we would like to explore more deeply because the data quality affects the model's quality.
We have built a training workflow that enables flexibility in labeling timestamps across the entire training audio to 1-second granularity.
While we succeeded in our baseline transfer learning experiment, having a human-labeled test dataset independent of the competition leaderboard would be helpful.
We can use ground-truth annotations to test how well an automated labeling process does at assigning primary, secondary, and no-call labels.

It would also be worth exploring human annotation in the sound separation process.
While we have listened to several examples to judge the quality of sound separation, we need a method to quantify quality.
In the case of sorting by the highest energy source, we may confuse a source with a significant amount of noise for the true birdcall just by the nature of higher entropy in the source.
In the case of sorting by the highest number of matching classifications from an existing model, we may need to produce a classifier that can distinguish between noise and birdcall for rare classes.
In this case, we continue to propagate uncertainty into the resulting labels, leading to poor performance, mainly when the number of examples is small.

We could introduce a metric to measure the resulting separation quality rigorously.
We could create many positive birdcall clips and have a human determine which channel mainly contains the primary species.
These clips let us see how well our automated channel selection process does at choosing the right channel based on human-generated labels.
However, this metric would not allow us to determine the separation quality in degenerate cases where a single separated source contains multiple bird vocalizations.

\subsection{Audio Source Separation}

The MixIT source separation model plays a significant role in the embedding pipeline used for our experiments.
One of the most significant benefits is noise suppression.
We also relied on sound-separated channels to generate training examples and semi-supervised labels.
Running an ablation study by evaluating the labeling process without access to the sound separation model would have been helpful.

We wanted to explore the 8-source model, which could have different performance characteristics than the 4-source model.
However, we decided to ignore this model because the training examples generally have few distinct vocalizations, and creating more source channels increases the necessary disk space for the intermediate files.

We are also interested in the effect of the sound separation model during inference.
However, we observed that the sound separation stage in training was a significant fraction of the computation budget.
There are also issues with different sample rates.
BirdNET expects audio sampled at 48kHz, while MixIT expects audio at 32KHz.
We did not attempt to integrate the model into this project because it would have put us over the submission time limit and required significant engineering effort to run inside the competition environment.

\subsection{Embedding Space and Transfer Learning}

Our experiments with interpolated embeddings in section \ref{sec:interpolated-emb model} had mixed results with respect to the baseline embedding model.
Including embedding context from other time intervals had substantially lower performance than our baseline.
The new labeling process may have overshadowed potential positive effects.
On the other hand, we saw a slight increase in our model performance when using the mean of pairs and triplets during model fitting.
Further research is necessary to determine whether it is valuable to manipulate embeddings similarly to mix up \cite{zhang2018mixup} that interpolates between training examples in the time domain to increase and augment training data.

We want to experiment with the embeddings from another model, such as the Google Research Bird Vocalization Classifier.
This classifier has seen more of the species in this competition than BirdNET.
It would be helpful to see how these two embeddings compare, and we could try this out as another feature.

Sequential models could be helpful in the competition by capturing dynamics and imbuing contextual information between embedding tokens.
The simplest model would be an auto-regressive linear model using an embedding from a single timestep to predict the next timestep optimized by a squared error loss.
We made initial forays into attention-based sequence-to-sequence models to address the output time-interval issue, but we needed more time to complete our experimentation.
Future work might explore data-driven methods like HAVOC \cite{brunton2017chaos} to analyze the dynamics of birdcall audio and their embeddings.

We would also like to explore the relationship between the sound-separated tracks and the embedding.
The separation model is constrained so that the sum of the sources results in the original track.
It would be interesting to verify a relationship between embeddings of various tracks by fitting a predictive model that takes tokens from each source to predict the embedding of the original track.
This line of thought does not directly help with model performance on the final task, but it does help understand the nature of the classifier embedding space.

\section{Conclusion}

In summary, our approach leveraged the embedding space learned by BirdNET to address the representation and labeling challenges in the competition. 
We developed a pipeline that included sound separation with MixIT, extraction of embedding tokens using BirdNET, and the creation of annotated datasets.
Our results showcased the competitive performance of our logistic regression baseline model as an effective method on unseen species and the comparative performance of various feature engineering work.
Our approach demonstrated the potential of transfer and semi-supervised learning for bird species classification in soundscapes.

\begin{acknowledgments}
Thanks to the Data Science at Georgia Tech (DS@GT) club for hosting our Kaggle competition team.
Thanks to DS@GT leadership for publicizing recruitment, particularly Krishi Manek as the Director of Projects.
Thanks to Erin Middlemas and Grant Williams for their support and engagement as initial team members.
\end{acknowledgments}

\bibliography{report.bib}

\begin{thebibliography}{13}
\expandafter\ifx\csname natexlab\endcsname\relax\def\natexlab#1{#1}\fi
\providecommand{\url}[1]{\texttt{#1}}
\providecommand{\href}[2]{#2}
\providecommand{\path}[1]{#1}
\providecommand{\DOIprefix}{doi:}
\providecommand{\ArXivprefix}{arXiv:}
\providecommand{\URLprefix}{URL: }
\providecommand{\Pubmedprefix}{pmid:}
\providecommand{\doi}[1]{\href{http://dx.doi.org/#1}{\path{#1}}}
\providecommand{\Pubmed}[1]{\href{pmid:#1}{\path{#1}}}
\providecommand{\bibinfo}[2]{#2}
\ifx\xfnm\relax \def\xfnm[#1]{\unskip,\space#1}\fi
%Type = Article
\bibitem[{Kahl et~al.(2023)Kahl, Denton, Klinck, Reers, Cherutich, Glotin,
  Go{\"e}au, Vellinga, Planqu{\'e}, and Joly}]{birdclef2023}
\bibinfo{author}{S.~Kahl}, \bibinfo{author}{T.~Denton},
  \bibinfo{author}{H.~Klinck}, \bibinfo{author}{H.~Reers},
  \bibinfo{author}{F.~Cherutich}, \bibinfo{author}{H.~Glotin},
  \bibinfo{author}{H.~Go{\"e}au}, \bibinfo{author}{W.-P. Vellinga},
  \bibinfo{author}{R.~Planqu{\'e}}, \bibinfo{author}{A.~Joly},
\newblock \bibinfo{title}{Overview of {BirdCLEF} 2023: Automated bird species
  identification in eastern africa},
\newblock \bibinfo{journal}{Working Notes of CLEF 2023 - Conference and Labs of
  the Evaluation Forum}  (\bibinfo{year}{2023}).
%Type = Inproceedings
\bibitem[{Joly et~al.(2023)Joly, Botella, Picek, Kahl, Go{\"e}au, Deneu,
  Marcos, Estopinan, Leblanc, Larcher, Chamidullin, \v{S}ulc, Hr\'{u}z,
  Servajean, Glotin, Planqué, Vellinga, Klinck, Denton, Eggel, Bonnet, and
  Müller}]{lifeclef2023}
\bibinfo{author}{A.~Joly}, \bibinfo{author}{C.~Botella},
  \bibinfo{author}{L.~Picek}, \bibinfo{author}{S.~Kahl},
  \bibinfo{author}{H.~Go{\"e}au}, \bibinfo{author}{B.~Deneu},
  \bibinfo{author}{D.~Marcos}, \bibinfo{author}{J.~Estopinan},
  \bibinfo{author}{C.~Leblanc}, \bibinfo{author}{T.~Larcher},
  \bibinfo{author}{R.~Chamidullin}, \bibinfo{author}{M.~\v{S}ulc},
  \bibinfo{author}{M.~Hr\'{u}z}, \bibinfo{author}{M.~Servajean},
  \bibinfo{author}{H.~Glotin}, \bibinfo{author}{R.~Planqué},
  \bibinfo{author}{W.-P. Vellinga}, \bibinfo{author}{H.~Klinck},
  \bibinfo{author}{T.~Denton}, \bibinfo{author}{I.~Eggel},
  \bibinfo{author}{P.~Bonnet}, \bibinfo{author}{H.~Müller},
\newblock \bibinfo{title}{{Overview of LifeCLEF 2023: evaluation of AI models
  for the identification and prediction of birds, plants, snakes and fungi}},
\newblock in: \bibinfo{booktitle}{International Conference of the
  Cross-Language Evaluation Forum for European Languages},
  \bibinfo{organization}{Springer}, \bibinfo{year}{2023}.
%Type = Article
\bibitem[{Kahl et~al.(2021)Kahl, Wood, Eibl, and Klinck}]{kahl2021birdnet}
\bibinfo{author}{S.~Kahl}, \bibinfo{author}{C.~M. Wood},
  \bibinfo{author}{M.~Eibl}, \bibinfo{author}{H.~Klinck},
\newblock \bibinfo{title}{Birdnet: A deep learning solution for avian diversity
  monitoring},
\newblock \bibinfo{journal}{Ecological Informatics} \bibinfo{volume}{61}
  (\bibinfo{year}{2021}) \bibinfo{pages}{101236}.
%Type = Misc
\bibitem[{McInnes et~al.(2020)McInnes, Healy, and Melville}]{mcinnes2020umap}
\bibinfo{author}{L.~McInnes}, \bibinfo{author}{J.~Healy},
  \bibinfo{author}{J.~Melville}, \bibinfo{title}{Umap: Uniform manifold
  approximation and projection for dimension reduction}, \bibinfo{year}{2020}.
  \href{http://arxiv.org/abs/1802.03426}{{\tt arXiv:1802.03426}}.
%Type = Inproceedings
\bibitem[{Denton et~al.(2022)Denton, Wisdom, and Hershey}]{denton2022improving}
\bibinfo{author}{T.~Denton}, \bibinfo{author}{S.~Wisdom},
  \bibinfo{author}{J.~R. Hershey},
\newblock \bibinfo{title}{Improving bird classification with unsupervised sound
  separation},
\newblock in: \bibinfo{booktitle}{ICASSP 2022-2022 IEEE International
  Conference on Acoustics, Speech and Signal Processing (ICASSP)},
  \bibinfo{organization}{IEEE}, \bibinfo{year}{2022}, pp.
  \bibinfo{pages}{636--640}.
%Type = Inproceedings
\bibitem[{Armbrust et~al.(2015)Armbrust, Xin, Lian, Huai, Liu, Bradley, Meng,
  Kaftan, Franklin, Ghodsi et~al.}]{armbrust2015spark}
\bibinfo{author}{M.~Armbrust}, \bibinfo{author}{R.~S. Xin},
  \bibinfo{author}{C.~Lian}, \bibinfo{author}{Y.~Huai},
  \bibinfo{author}{D.~Liu}, \bibinfo{author}{J.~K. Bradley},
  \bibinfo{author}{X.~Meng}, \bibinfo{author}{T.~Kaftan},
  \bibinfo{author}{M.~J. Franklin}, \bibinfo{author}{A.~Ghodsi}, et~al.,
\newblock \bibinfo{title}{Spark sql: Relational data processing in spark},
\newblock in: \bibinfo{booktitle}{Proceedings of the 2015 ACM SIGMOD
  international conference on management of data}, \bibinfo{year}{2015}, pp.
  \bibinfo{pages}{1383--1394}.
%Type = Misc
\bibitem[{Lui(2023)}]{Luigi2023}
\bibinfo{title}{Luigi 2.8.13 documentation},
  \bibinfo{howpublished}{\url{https://luigi.readthedocs.io/en/stable/}},
  \bibinfo{year}{2023}. \bibinfo{note}{Accessed: 2023-06-07}.
%Type = Article
\bibitem[{Pedregosa et~al.(2011)Pedregosa, Varoquaux, Gramfort, Michel,
  Thirion, Grisel, Blondel, Prettenhofer, Weiss, Dubourg, Vanderplas, Passos,
  Cournapeau, Brucher, Perrot, and Duchesnay}]{scikit-learn}
\bibinfo{author}{F.~Pedregosa}, \bibinfo{author}{G.~Varoquaux},
  \bibinfo{author}{A.~Gramfort}, \bibinfo{author}{V.~Michel},
  \bibinfo{author}{B.~Thirion}, \bibinfo{author}{O.~Grisel},
  \bibinfo{author}{M.~Blondel}, \bibinfo{author}{P.~Prettenhofer},
  \bibinfo{author}{R.~Weiss}, \bibinfo{author}{V.~Dubourg},
  \bibinfo{author}{J.~Vanderplas}, \bibinfo{author}{A.~Passos},
  \bibinfo{author}{D.~Cournapeau}, \bibinfo{author}{M.~Brucher},
  \bibinfo{author}{M.~Perrot}, \bibinfo{author}{E.~Duchesnay},
\newblock \bibinfo{title}{Scikit-learn: Machine learning in {P}ython},
\newblock \bibinfo{journal}{Journal of Machine Learning Research}
  \bibinfo{volume}{12} (\bibinfo{year}{2011}) \bibinfo{pages}{2825--2830}.
%Type = Inproceedings
\bibitem[{Chen and Guestrin(2016)}]{chen2016xgboost}
\bibinfo{author}{T.~Chen}, \bibinfo{author}{C.~Guestrin},
\newblock \bibinfo{title}{{XGBoost}: A scalable tree boosting system},
\newblock in: \bibinfo{booktitle}{Proceedings of the 22nd ACM SIGKDD
  International Conference on Knowledge Discovery and Data Mining}, KDD '16,
  \bibinfo{publisher}{ACM}, \bibinfo{address}{New York, NY, USA},
  \bibinfo{year}{2016}, pp. \bibinfo{pages}{785--794}. \URLprefix
  \url{http://doi.acm.org/10.1145/2939672.2939785}.
  \DOIprefix\doi{10.1145/2939672.2939785}.
%Type = Misc
\bibitem[{Head et~al.(2018)Head, MechCoder, Louppe, Shcherbatyi, fcharras,
  Vinícius, cmmalone, Schröder, nel215, Campos, Young, Cereda, Fan, rene rex,
  Shi, Schwabedal, carlosdanielcsantos, Hvass-Labs, Pak, SoManyUsernamesTaken,
  Callaway, Estève, Besson, Cherti, Pfannschmidt, Linzberger, Cauet, Gut,
  Mueller, and Fabisch}]{head2018skopt}
\bibinfo{author}{T.~Head}, \bibinfo{author}{MechCoder},
  \bibinfo{author}{G.~Louppe}, \bibinfo{author}{I.~Shcherbatyi},
  \bibinfo{author}{fcharras}, \bibinfo{author}{Z.~Vinícius},
  \bibinfo{author}{cmmalone}, \bibinfo{author}{C.~Schröder},
  \bibinfo{author}{nel215}, \bibinfo{author}{N.~Campos},
  \bibinfo{author}{T.~Young}, \bibinfo{author}{S.~Cereda},
  \bibinfo{author}{T.~Fan}, \bibinfo{author}{rene rex}, \bibinfo{author}{K.~K.
  Shi}, \bibinfo{author}{J.~Schwabedal}, \bibinfo{author}{carlosdanielcsantos},
  \bibinfo{author}{Hvass-Labs}, \bibinfo{author}{M.~Pak},
  \bibinfo{author}{SoManyUsernamesTaken}, \bibinfo{author}{F.~Callaway},
  \bibinfo{author}{L.~Estève}, \bibinfo{author}{L.~Besson},
  \bibinfo{author}{M.~Cherti}, \bibinfo{author}{K.~Pfannschmidt},
  \bibinfo{author}{F.~Linzberger}, \bibinfo{author}{C.~Cauet},
  \bibinfo{author}{A.~Gut}, \bibinfo{author}{A.~Mueller},
  \bibinfo{author}{A.~Fabisch},
  \bibinfo{title}{scikit-optimize/scikit-optimize: v0.5.2},
  \bibinfo{year}{2018}. \URLprefix
  \url{https://doi.org/10.5281/zenodo.1207017}.
  \DOIprefix\doi{10.5281/zenodo.1207017}.
%Type = Article
\bibitem[{Stowell and Plumbley(2013)}]{stowell2013open}
\bibinfo{author}{D.~Stowell}, \bibinfo{author}{M.~D. Plumbley},
\newblock \bibinfo{title}{An open dataset for research on audio field recording
  archives: freefield1010},
\newblock \bibinfo{journal}{arXiv preprint arXiv:1309.5275}
  (\bibinfo{year}{2013}).
%Type = Misc
\bibitem[{Zhang et~al.(2018)Zhang, Cisse, Dauphin, and
  Lopez-Paz}]{zhang2018mixup}
\bibinfo{author}{H.~Zhang}, \bibinfo{author}{M.~Cisse}, \bibinfo{author}{Y.~N.
  Dauphin}, \bibinfo{author}{D.~Lopez-Paz}, \bibinfo{title}{mixup: Beyond
  empirical risk minimization}, \bibinfo{year}{2018}.
  \href{http://arxiv.org/abs/1710.09412}{{\tt arXiv:1710.09412}}.
%Type = Article
\bibitem[{Brunton et~al.(2017)Brunton, Brunton, Proctor, Kaiser, and
  Kutz}]{brunton2017chaos}
\bibinfo{author}{S.~L. Brunton}, \bibinfo{author}{B.~W. Brunton},
  \bibinfo{author}{J.~L. Proctor}, \bibinfo{author}{E.~Kaiser},
  \bibinfo{author}{J.~N. Kutz},
\newblock \bibinfo{title}{Chaos as an intermittently forced linear system},
\newblock \bibinfo{journal}{Nature communications} \bibinfo{volume}{8}
  (\bibinfo{year}{2017}) \bibinfo{pages}{19}.

\end{thebibliography}
\end{document}